\documentclass[journal=ancac3,manuscript=article]{achemso}

\usepackage[version=3]{mhchem} 

\usepackage[utf8]{inputenc}
\usepackage[T1]{fontenc}
\usepackage{lmodern}
\usepackage{bm}
\usepackage{comment}

\usepackage{nicefrac}

\usepackage{siunitx}
\usepackage{amsmath}
\usepackage{color}

\author{Xiaofei Xiao}
\altaffiliation{Contributed equally to this work}
\email{xiaofei.xiao15@imperial.ac.uk}
\author{Tingjun Zheng}
\altaffiliation{Contributed equally to this work}
\author{Alex Vanstone}
\altaffiliation{Contributed equally to this work}

\author{Daniel S. Bromley}

\author{Holly Holder}

\author{Kilian D. Stenning}

\author{Jack C. Gartside}

\author{Will R. Branford}

\author{Rupert F. Oulton}
\affiliation[ICL]
{Blackett Laboratory, Imperial College London, London SW7 2AZ, United Kingdom}
\email{r.oulton@imperial.ac.uk}

\title[An \textsf{achemso} demo]
  {Enhanced Magneto-Optical Kerr Effect in Nano-Magnetic Metamaterials with Near-Perfect Coherent Absorption}

\abbreviations{MOKE, SEM, FTIR, ASI, FDTD, Py, PMMA, MIBK, IPA}
\keywords{metamaterials, optical absorption enhancement, magneto-optical Kerr effect, nanomagnetism, all-optical magnetic switching,  magnetic light absorption}

\begin{document}

\begin{tocentry}

\includegraphics[width=0.93\textwidth]{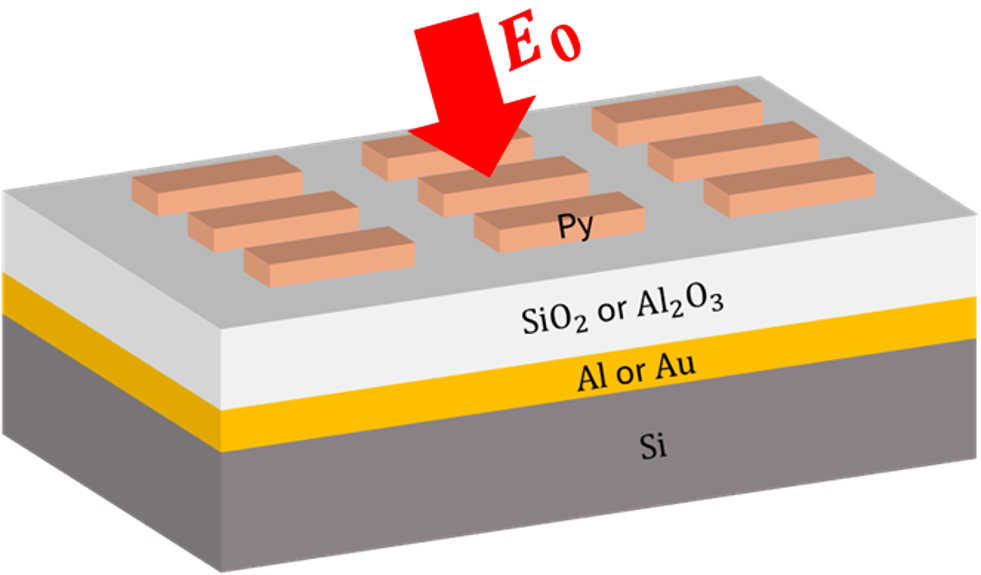}





\end{tocentry}

\begin{abstract}
  The ability to control magnetic systems without external magnetic fields has attracted attention recently, with all-optical magnetic switching showing increasing promise. Gaining fine control over light-matter interactions in magnetic nanostructures is essential for advancing opto-magnetic technologies, but remains challenging because strong optical coupling must be achieved within ultrathin magnetic structures at the target wavelength. Here, we show that coherent optical interference can increase this interaction in patterned magnetic metasurfaces. Arrays of permalloy nanomagnets are placed on a thin dielectric spacer above a reflective back-plane, forming a magnetic metasurface whose reflected wavefront is defined by the nanomagnet geometry, fill factor, and spacer thickness. Destructive interference between the metasurface and back-plane reflections yields near-perfect absorption, exceeding 95\% for circular nanomagnet arrays in both simulation and experiment. The absorbed power is dissipated almost entirely within the nanomagnets, with only a few percent lost in the metallic back-plane. Enhanced absorption is also accompanied by a stronger magneto-optical Kerr effect, driven by both increased light–nanomagnet interaction and reduced reflected background. These results provide a general approach for achieving high optical absorption and strong magneto-optical response in nanoscale magnetic systems, supporting advances in optically addressable spintronics, magnetic sensing, and all-optical magnetic control.
\end{abstract}

\section{Introduction}

The use of light to manipulate magnetic states has come to prominence in recent years through investigations of optical demagnetization \cite{zhang2000laser,pfau2012ultrafast,eschenlohr2013ultrafast}, all-optical switching \cite{kirilyuk2010ultrafast,kimel2019writing,maccaferri2020nanoscale,stenning2023low,holder2026magnetic}, optically addressed magnetic sensors \cite{chen2023review}, and plasmon-assisted thermal control of nanomagnets \cite{pancaldi2019selective,gypens2022thermoplasmonic}, with applications in spintronic computing and data storage. Enhancing light's interaction with magnetic materials has thus become important for a large range of emerging technologies. The most widely used magnetic materials are metallic, and their loss is strongly frequency dependent \cite{song2021inorganic}; many ferromagnetic alloys absorb strongly in particular spectral windows. The difficulty is achieving strong interaction at the target wavelength. Magnetic device designs require layer thicknesses from a few nanometers to tens of nanometers, which substantially limits the optical interaction volume, especially when patterned into nanostructures. This has motivated interest in the plasmonics of magnetic metals \cite{maksymov2016magneto,yang2022observation}, however, the same large imaginary permittivity that provides absorption also damps the resonance, limiting plasmonic enhancement. Indeed, hybrid plasmonic systems that combine magnetic and noble metals have also been considered \cite{schulz2024roadmap}. One strategy to overcome this limitation is to use thin-film interference, where multi-layer stacks of dielectrics and metals \cite{kats2012ultra,hendriks2021enhancing} manipulate light waves to control their reflection, transmission and absorption. For example, large enhancements of the magneto-optical Kerr effect (MOKE) were reported in planar stacks of dielectric and magnetic layers by enhancing absorption \cite{kim2020extreme}. Enhanced absorption has also recently been shown to be important in all-optical switching of states in magnetic metamaterials comprising dense arrays of strongly coupled nanomagnets (artificial spin ice - ASI) using low power continuous wave lasers \cite{stenning2023low}.

Achieving high absorption in thin materials is essential for a variety of technologies throughout the electromagnetic spectrum \cite{hagglund2010maximized,stenning2023low,kats2016optical,rider2019perspective,xiao2024nanofocusing,schulz2024roadmap,fischer2024surface,shautsova2018plasmon}, including photovoltaic and photochemical cells, photodetectors, optical filters, stealth technology, and thermal light sources. Conventional routes to increasing interaction and absorption rely on techniques, such as thin-film interference, plasmonics and metamaterials \cite{abdulkarim2022review,yu2019broadband,hedayati2014review,zheludev2012metamaterials,yu2011light,baranov2015perfect,xiao2018theoretical,xiao2021ultrabroad}. Researchers have also explored the use of thin but absorptive materials (including semiconductors and metals) as nanometer-thick optical coatings \cite{kats2016optical,kats2013nanometre} to control reflection and transmission across the ultraviolet, visible, infrared, and terahertz spectral ranges \cite{dotan2013resonant,park2014omnidirectional,zhang2014refractive,streyer2013strong,kats2012ultra}. While thin-film interference has been shown to be effective in continuous 
planar magnetic films~\cite{kim2020extreme}, enhancing light's interaction 
with magnetic nanoscale structures, such as ASI~\cite{gartside2022reconfigurable}, 
warrants further exploration. Existing routes to enhancing magneto-optical 
Kerr effect (MOKE) responses include interference-based cavities and 
multilayer structures~\cite{nyvlt1996magneto}, plasmonic and magnetoplasmonic 
nanostructures that concentrate near fields around magnetic 
elements~\cite{belotelov2011enhanced,armelles2013magnetoplasmonics,lopez2020enhanced,wang2025magnetoplasmonic}, 
all-dielectric metasurfaces and photonic resonators exploiting Mie, guided, 
or quasi-bound states in the continuum 
resonances~\cite{barsukova2017magneto,gao2023optimizing}, intrinsic and 
interfacial material engineering~\cite{gu2017enhanced}, and measurement-side 
amplification schemes such as weak-value amplification~\cite{huang2023weak}. 
Each of these entails trade-offs in Ohmic loss, fabrication and angular 
tolerance, tunability, or in whether the intrinsic light-magnetization 
interaction is enhanced at all; a detailed comparison is provided in the 
Supporting Information.

In this report, we demonstrate enhanced light absorption in permalloy 
nanomagnet arrays constructed atop a thin-dielectric layer and highly 
reflective back-plane. The array operates as an optical magnetic metasurface, 
where the permalloy thickness, fill factor, and nanomagnet shape control 
reflection and transmission. Through this control, we are able to establish 
destructive interference between reflections from the permalloy metasurface 
and the highly reflective back-plane, leading to near-perfect coherent 
absorption in the nanomagnet array exceeding 95\% in total, with only a few percent of the incident power dissipated in the metallic back-plane. We also report enhanced 
MOKE in nanomagnet arrays, which arises as a combination of the enhanced 
light-nanomagnet interaction as well as their suppressed reflectivity. 
This interference-assisted absorption mechanism requires neither extreme 
plasmonic hot-spot confinement nor ultra-narrow resonance linewidths, and 
remains lithographically tunable through nanomagnet geometry alone.

\begin{figure}
    \centering
    \includegraphics[width=1\textwidth]{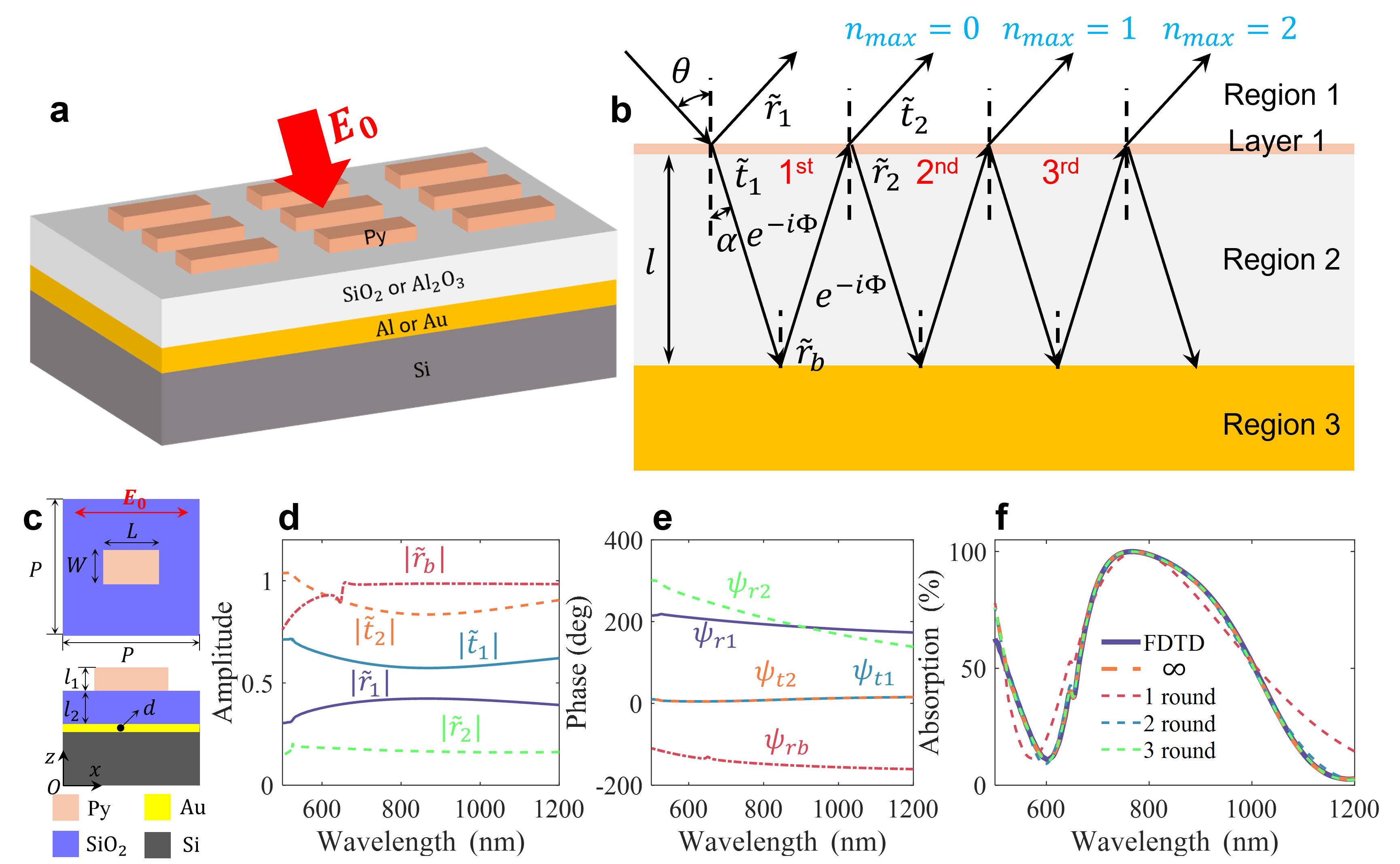}
    \caption{\textbf{Schematic of structured nanometre optical coatings.} (\textbf{a}), Schematic illustration of isolated nanostructured lossy material (Py nanomagnets) patterned on a metallic film (Al or Au), deposited on a thick substrate (Si) with a dielectric coating (SiO$_2$ or Al$_2$O$_3$). (\textbf{b}), Schematic of a simplified four-layer structure under oblique incidence. An incident beam from Region 1 (air) is partially reflected at Layer 1 (Py), which is assumed to have negligible thickness, and partially transmitted into Region 2, which has a finite thickness $l$. The transmitted wave propagates through Region 2 until it reaches Region 3 (the metallic ground plane), accumulating a complex propagation phase. Upon reflection at the ground plane and acquisition of an additional propagation phase, the wave undergoes partial reflection at the interface between Region 2 and Layer 1, and partial transmission back into Region 1 via Layer 1. This process repeats indefinitely, with the light gradually diminishing. (\textbf{c}), Schematic of the unit cell in an exemplary nanomagnet system. The rectangular permalloy nanomagnet arrays are arranged in a square lattice. The SiO$_2$ and the gold mirror serve as a cavity system.  Amplitude (\textbf{d}) and phase (\textbf{e}) of the reflection and transmission coefficients in the theory for an illustrative system: The designed nanomagnet system has dimensions: $l_1$ = 20 nm, $l_2$ = 375 nm, $L$ = 250 nm, $W$ = 100 nm, $d$ = 250 nm, and $P$ = 360 nm. The incident light is x-polarized. (\textbf{f}), Absorption spectra were calculated using both the interference model (dashed lines) and a comprehensive simulation (solid line) of the entire system. The legends denote the round of reflection employed in the interference theory, where $n$ rounds of reflection correspond to $n_{\mathrm{max}} = n-1$ in Equation 1 ($n_{\mathrm{max}} = \infty$, 0, 1, and 2).}
    \label{Fig:Schematic}
\end{figure}

\section{Results and discussion}
\subsection{System Description}

The permalloy (Py) metasurface interference system consists of four layers, as shown in Figure \ref{Fig:Schematic}a. Light is incident from the top, considered air, onto the metasurface layer of the permalloy (Ni$_{81}$Fe$_{19}$), whose structure consists of circular disks or bars. As shown in Figure \ref{Fig:Schematic}b, this layer can be viewed as an effective medium of zero thickness, where its true thickness, fill factor, and nanomagnet shape determine the reflection coefficients from Region 1 (2) to Region 1 (2), $\tilde{r}_1= |\tilde{r}_1|e^{i\psi_{r1}}$ and $\tilde{r}_2= |\tilde{r}_2|e^{i\psi_{r2}}$, and the transmission coefficients from Region 1 (2) to Region 2 (1), $\tilde{t}_1= |\tilde{t}_1|e^{i\psi_{t1}}$ and $\tilde{t}_2= |\tilde{t}_2|e^{i\psi_{t2}}$. 
Reflection and transmission from the metasurface are assumed to be specular with no diffraction orders, since the nanomagnets and their period are both subwavelength in scale. The permalloy metasurface sits on a dielectric layer of variable thickness, $l$, which controls the phase change, $\Phi=k_zl$, of light passing through it, where $k_z$ is the z-component of the wavevector. The final layer is the metal back-plane (either aluminum or gold, in our case), which should be highly reflective, $\tilde{r}_b = |\tilde{r}_b|e^{i\psi_{rb}}$, across the desired operation wavelengths; in this case, the visible and near-infrared. Note that the reflection phase of the metal back-plane $\psi_{rb}$ is not simply $\pi$ due to the finite conductivity of real metals and their highly dispersive permittivity in the frequency region of interest \cite{kats2012ultra}. The near-perfect absorption in this system originates from Fabry--Pérot-type interference induced by multiple reflections within the multilayer structure, enabling coherent absorption. The total reflection, $\tilde{r}_t$, accounting for multiple reflections within the system is,
\begin{equation}
    \tilde{r}_{t} = \tilde{r}_{1} + \tilde{t}_{1}  \tilde{r}_{b} \tilde{t}_{2} e^{2i\Phi} \sum_{n=0}^{n_{\textnormal{max}}\to\infty} (\tilde{r}_{2}  \tilde{r}_{b} e^{2i\Phi})^n = \tilde{r}_{1} + \frac{\tilde{t}_{1} \tilde{t}_{2}  \tilde{r}_{b} e^{2i\Phi} }{1 - \tilde{r}_{2} \tilde{r}_{b} e^{2i\Phi}}.
    \label{Eq:totalrefelctionCoffs}
\end{equation}

The total absorption of the system can be calculated by $A_t(\omega) = 1 - R_t(\omega) - T_t(\omega) \approx 1 - |r_t|^2$, since $T_t(\omega) =0$ for a thick, highly reflective back-plane. For the aluminum back-plane, the ground plane absorbs no more than a few percent of the incident light across the wavelength range studied here (see Supporting Information, Figure S1), while the rest is absorbed by the magnetic metasurface; the same holds for gold above its interband edge. Our goal is thus to suppress the reflection, \( R_t \), thereby enhancing absorption by the metasurface. 

To analyze the physics, consider the exemplary nanomagnet system shown in Figure~\ref{Fig:Schematic}c. We first computed the reflection and transmission coefficients with their respective phases for the magnetic metasurface alone for normal incidence as shown in Figures~\ref{Fig:Schematic}d and e. Using these effective parameters, the normal-incidence absorption of the complete structure with metal back-plane was then calculated using Equation~\ref{Eq:totalrefelctionCoffs}, with varying numbers of round-trip reflections, \( n_{max} \), as shown in Figure~\ref{Fig:Schematic}f. The absorption of the complete optical coating system was also rigorously simulated as a three-dimensional (3D) coupled magnetic metamaterial multi-layer structure using the finite-difference time-domain (FDTD) method (Lumerical FDTD). The computed absorption obtained from the simplified interference model agrees closely with the rigorous numerical simulations. For decreased round-trip number, the agreement deteriorates only marginally, with \( n_{max} =2 \) already yielding negligible deviation. When \( n_{max}=0 \), the interference model still performs sufficiently well and can be simplified to,
\begin{equation}
\tilde{r}_{t} \approx \tilde{r}_{1} + \tilde{t}_{1} \tilde{r}_{b} \tilde{t}_{2} e^{2i\Phi} \approx  |\tilde{t}_{1}| |\tilde{t}_{2}| |\tilde{r}_b|e^{i(\psi_{rb} + 2\Phi)}-|\tilde{r}_1|.
\end{equation}
where we have applied the values \(\psi_{r1} \approx \pi\), \(\psi_{t1} = \psi_{t2} \approx 0\), from the computed reflection and transmission coefficients of the magnetic layers, shown in Fig~\ref{Fig:Schematic}d and e. To enhance absorption, destructive interference of the reflected waves is necessary, where the SiO$_2$ layer thickness should be chosen such that \(\psi_{rb} + 2\Phi \approx 2m \pi\) for an integer \( m \). Near-perfect absorption can then be achieved when $|\tilde{t}_{1}| |\tilde{t}_{2}| |\tilde{r}_b|=|\tilde{r}_1|$, which can be controlled by varying the permalloy nanostructure shape as well as the fill factor. 

In the remainder of the paper, our quantitative predictions use FDTD simulations to calculate the optical absorption of the system. In our experiments, we study the absorption by the permalloy nanomagnet metasurface using Fourier transform infrared (FTIR) spectroscopy, while we study the magneto-optical interaction using longitudinal MOKE measurements. For normal-incidence simulations and measurements, the polarization is defined with respect to the Cartesian coordinates, where \(x\)- and \(y\)-polarized light correspond to electric fields oriented along the \(x\)- and \(y\)-axes, respectively. For oblique-incidence MOKE measurements, the polarization is defined relative to the plane of incidence. Since the measurements were performed in the \(xz\)-plane, \(s\)-polarization corresponds to an electric field along the \(y\)-direction, while \(p\)-polarization corresponds to an electric field within the \(xz\)-plane.

\subsection{Tuning absorption via reflection phase}

\begin{figure}
    \centering
    \includegraphics[width=0.7\textwidth]{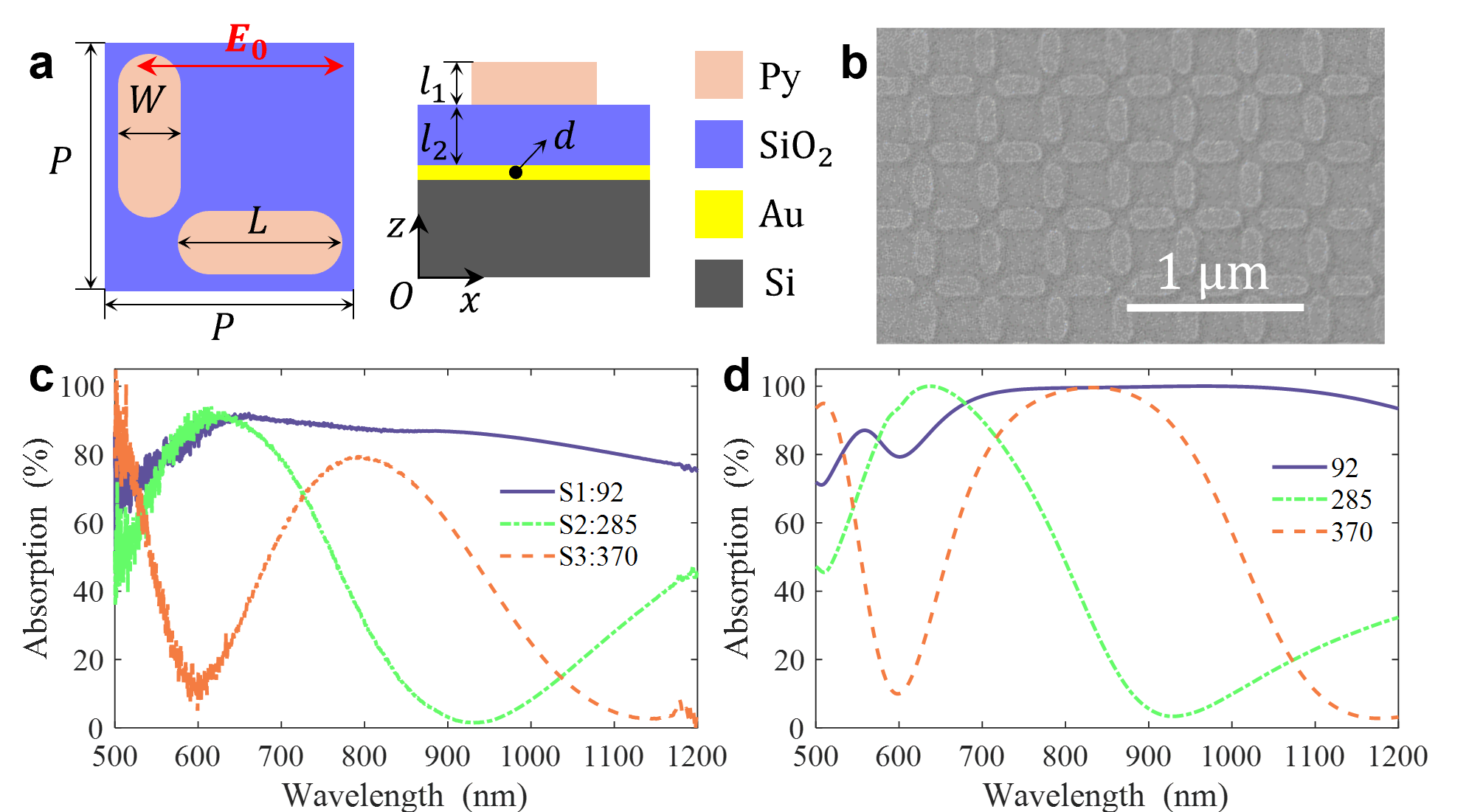}
    \caption{\textbf{Thickness-controlled absorption of the structured nanometre optical coatings.} (\textbf{a}), Schematic of the nanomagnet array structures with varying cavity thickness ($l_2$). (\textbf{b}), SEM image of the fabricated periodic nanomagnet structures. The scale bar represents $\SI{1}{\micro\meter}$. Measured (\textbf{c}) and simulated (\textbf{d}) absorption spectra of nanomagnet arrays for varying SiO$_2$ thickness ($l_2$). In (\textbf{c}), S1-S3 label the three fabricated samples, followed by their nominal cavity thickness $l_2$ in nm; the simulations in (\textbf{d}) use the same nominal thicknesses. The designed nanomagnet system has dimensions: $l_1$ = 20 nm, $L$ = 250 nm, $W$ = 110 nm, $d$ = 250 nm, and $P$ = 360 nm. The incident light is x-polarized.}
    \label{Fig:Thickness_FTIRAndsimulation}
\end{figure}

We first explore the enhancement of optical absorption by tuning the thickness $l$ of the dielectric spacer layer. Figure \ref{Fig:Thickness_FTIRAndsimulation}a schematically depicts the magnetic metasurface unit cell and layer structure. The metasurface consists of permalloy nanobars arranged in a square ASI geometry, where half of the bars are oriented along the x-axis and the other half along the y-axis. A representative scanning electron microscope (SEM) image of the fabricated structures is shown in Figure~\ref{Fig:Thickness_FTIRAndsimulation}b, demonstrating high-quality patterning over large areas. The dielectric spacer thickness is varied to modulate the optical cavity properties and achieve spectral tuning of the absorption.

FTIR measurements were performed at near-normal incidence over the 500--1200 nm spectral range for varying silica thicknesses, as shown in Figure~\ref{Fig:Thickness_FTIRAndsimulation}c. The corresponding FDTD simulations, shown in Figure~\ref{Fig:Thickness_FTIRAndsimulation}d, reproduce the measured absorption spectra well. Minor discrepancies are attributed to fabrication tolerances, such as edge roughness, grain boundaries, and slight variations in silica thickness. The measured absorption of this ASI nanobar array peaks at $>80\%$ and can be spectrally tuned across a wide range of frequencies by modifying the spacer thickness, while simulations for the same structure give $>95\%$. Nonetheless, the correspondence of experiment and theory is close with spectral features clearly correlated. Notably, an octave-spanning broadband absorption is obtained for the lowest order of spacer thickness.

The 500--1200\,nm window is set by the materials and by our instrument. Gold loses
its near-unity reflectivity below its interband edge near 515\,nm (see Supporting
Information, Figure~S3a,b), degrading the interference condition for the Au
back-plane of Figure~\ref{Fig:Thickness_FTIRAndsimulation}; aluminum retains high reflectivity further into the blue,
so for the remaining samples the short-wavelength bound is instead set by the
usable range of our FTIR system. At long wavelengths the round-trip phase detunes
from the value designed at 633\,nm. Neither bound is fundamental, since the
condition can be re-established elsewhere by scaling $l_2 \propto \lambda$ and
choosing a mirror that remains reflective there.

\subsection{Tuning absorption via magnetic metasurface fill-factor}

\begin{figure}[h]
    \centering
    \includegraphics[width=0.7\textwidth]{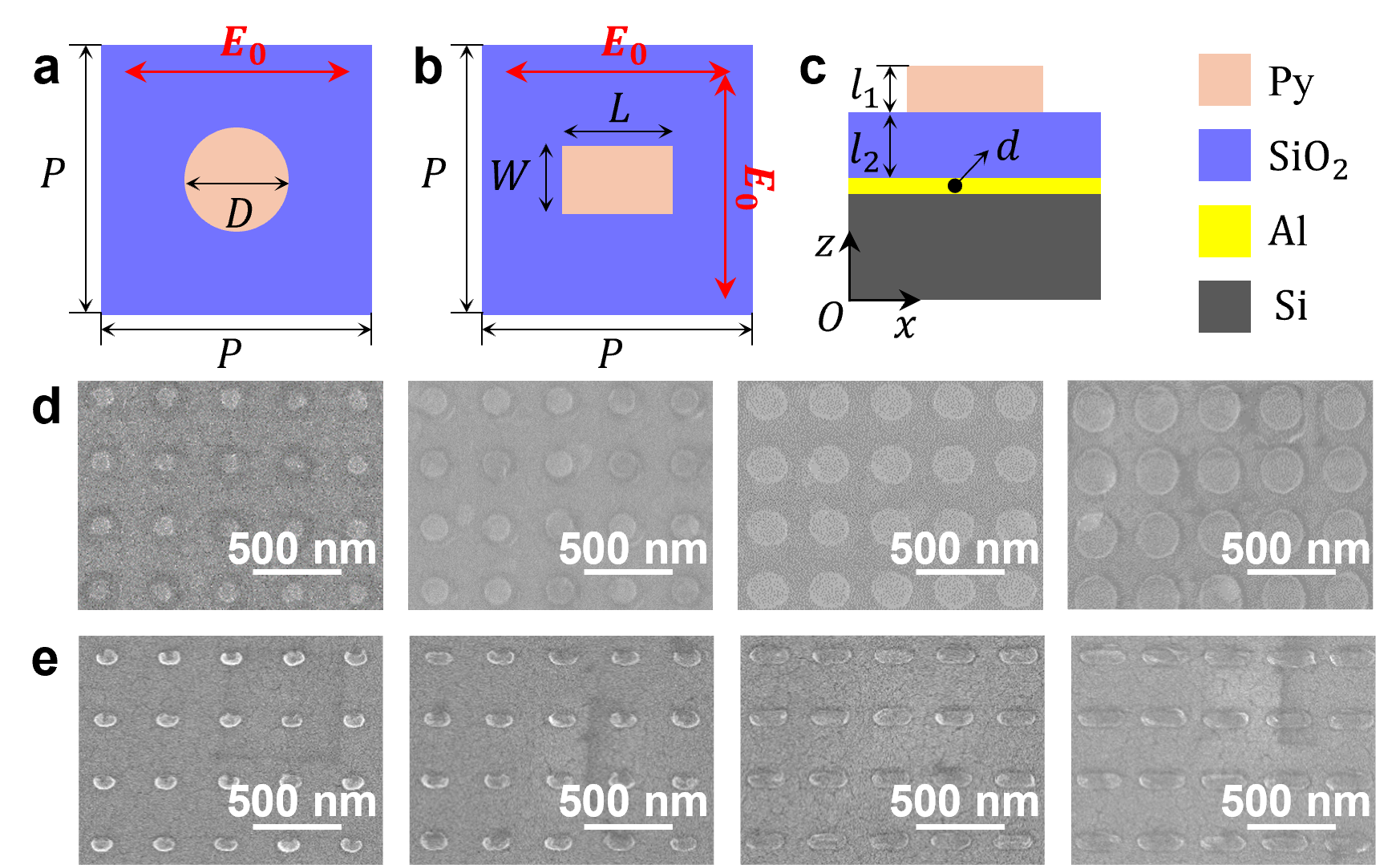}
    \caption{\textbf{Structure geometry and fabricated samples for fill-factor-controlled absorption.} (\textbf{a},\textbf{b}) Top view and (\textbf{c}) side view of the unit cell of the nanomagnet system. The (\textbf{a}) circular or (\textbf{b}) rectangular permalloy nanomagnet arrays are arranged in a square lattice. The SiO$_2$ and the aluminium mirror serve as a cavity system. The absorption response can be controlled by varying the size of the permalloy nanomagnet, the thickness of the SiO$_2$ ($l_2$) and the nanomagnets ($l_1$). The nominal geometric parameters are $l_1$ = 20 nm, $l_2$ = 92 nm, $W$ = 100 nm, and $d$ = 250 nm. The period of the square lattice is fixed to $P$ = 360 nm. All structures were fabricated on a silicon substrate. Linearly polarized light at normal incidence illuminates the system from the air side. For circular nanomagnets, we use x-polarized light, while for rectangular nanomagnets, we apply both x- and y-polarized light. SEM images of the fabricated (\textbf{d}) circular (of diameter $D$) and (\textbf{e}) rectangular (of length $L$ and width $W$) nanomagnets with various sizes. The scale bars represent 500 nm.}
    \label{Figure:SEM_circlular_bar}
\end{figure}

We now examine absorption control via the metasurface reflection and transmission using both the shape and fill factor of magnetic nanostructures. In ASI structures, the optical coupling between perpendicular permalloy bars complicates our analysis, so we use new metasurfaces with single nano-structure bases. Figure \ref{Figure:SEM_circlular_bar} presents the two sets of nanomagnet arrays used. Figures~\ref{Figure:SEM_circlular_bar}a and b show circular and rectangular nanomagnet unit cells, respectively, with the arrows indicating the incident light polarization used in each case. Absorption tuning is achieved by adjusting the nanomagnet dimensions; the diameter $D$ in the circular case, and the length $L$ and width $W$ in the rectangular case. These parameters thus also control the fill-factor. Permalloy is employed as the magnetic medium, while SiO\textsubscript{2} serves as the lossless dielectric spacer layer, fixed at a thickness of $92$ nm in this study to provide broad absorption as observed in Figure ~\ref{Fig:Thickness_FTIRAndsimulation}. Figure ~\ref{Figure:SEM_circlular_bar}c shows the layers of the design. Representative SEM images of the fabricated circular and rectangular arrays are shown in Figures~\ref{Figure:SEM_circlular_bar}d and e, respectively, highlighting the high fidelity of the nanofabrication.

\begin{figure}
    \centering
    \includegraphics[width=1\textwidth]{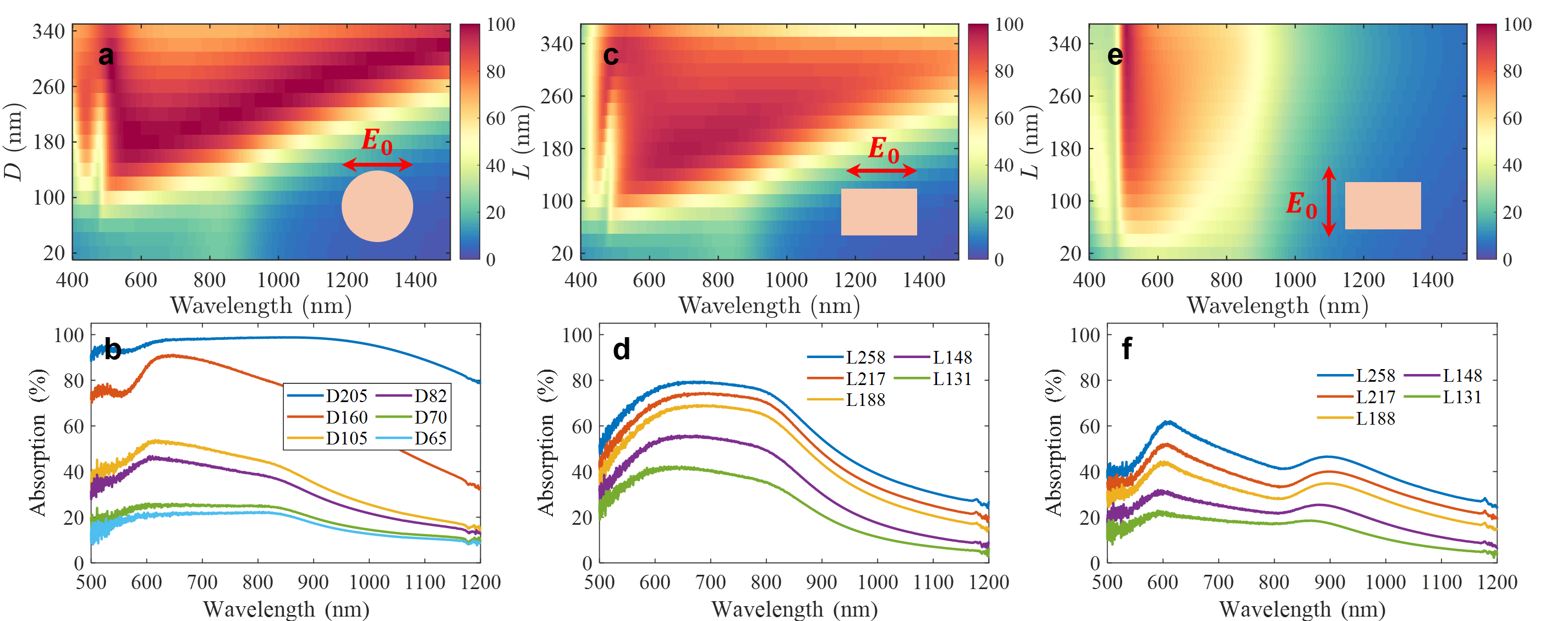}
    \caption{\textbf{Fill-factor-controlled absorption in structured nanometre optical coatings.} Simulated (\textbf{a}) and measured (\textbf{b}) absorption spectra of circular nanomagnet arrays with different diameters $D$ under x-polarized incidence. Simulated (\textbf{c}) and measured (\textbf{d}) absorption spectra of rectangular nanomagnet arrays with different lengths $L$ under x-polarized incidence. Simulated (\textbf{e}) and measured (\textbf{f}) absorption spectra of rectangular nanomagnet arrays with different lengths $L$ under y-polarized incidence. Details of the geometric parameters are provided in Figure~\ref{Figure:SEM_circlular_bar}.
    }
    \label{Figure:circlular_bar_FTIR}
\end{figure}

We fabricated six samples with circular nanomagnets of varying nominal diameters and characterized their optical response using FTIR spectroscopy under linearly polarized light (x-polarization). The simulated and experimentally measured absorption spectra are presented in Figures~\ref{Figure:circlular_bar_FTIR}a and~\ref{Figure:circlular_bar_FTIR}b, respectively. The two sets of results show excellent agreement, validating our simple and intuitive interference model. Notably, in the simulations, the absorption is enhanced at the surface plasmon resonance of the permalloy nanostructure, identified by the peak absorption red-shifting approximately linearly with increasing diameter. This condition corresponds to $|\tilde{t}_{1}| |\tilde{t}_{2}| |\tilde{r}_b|\approx|\tilde{r}_1|$, where the reflection is strongly suppressed. For the circular-disk metasurface, the absorption exceeds 95\% at $D=205$ nm in both simulation and experiment, as shown in Figure~\ref{Figure:circlular_bar_FTIR}b, corresponding to a fill factor of 25.5\%. This shows that near-perfect destructive interference of the reflection may be achieved, and thus near perfect absorption in the permalloy nanostructures.

Figure~\ref{Figure:circlular_bar_FTIR}c--f present the simulated and measured absorption spectra of the rectangular bar configuration for x-polarized incidence (Figure~\ref{Figure:circlular_bar_FTIR}c,d) and y-polarized incidence (Figure~\ref{Figure:circlular_bar_FTIR}e,f). Similar behavior to the circular configuration is observed, with resonant absorption influenced by the surface plasmon resonance of the bar. Owing to the relatively weak surface plasmon resonance of permalloy nanostructures, the fill factor plays a dominant role in determining the peak absorption. For the best-performing geometry in this series, the experimental peak absorption approaches $80\%$, for a fill factor of 19.9\%. In the case of incident polarization perpendicular to the bars, significant absorption may also occur, unless the bar width is reduced. Reducing the bar width reduces both fill-factor and the surface plasmon resonance of the short axis of the bar.

It should be noted that the present work examines and tunes the light-nanomagnet interaction within a reduced parameter space, primarily through the Fabry--P\'erot interference condition and the geometrical fill factor. Other nanomagnet parameters are kept fixed because they strongly influence the magnetic response, allowing the optical effects to be isolated. A more comprehensive optimization incorporating additional degrees of freedom, such as nanomagnet thickness, spacing, morphology, and material properties, may further enhance the interaction and could be explored in future work using inverse-design or machine-learning approaches. 

\subsection{Enhanced MOKE in Ferromagnetic Nanobars}

\begin{figure}
    \centering
    \includegraphics[width=0.7\textwidth]{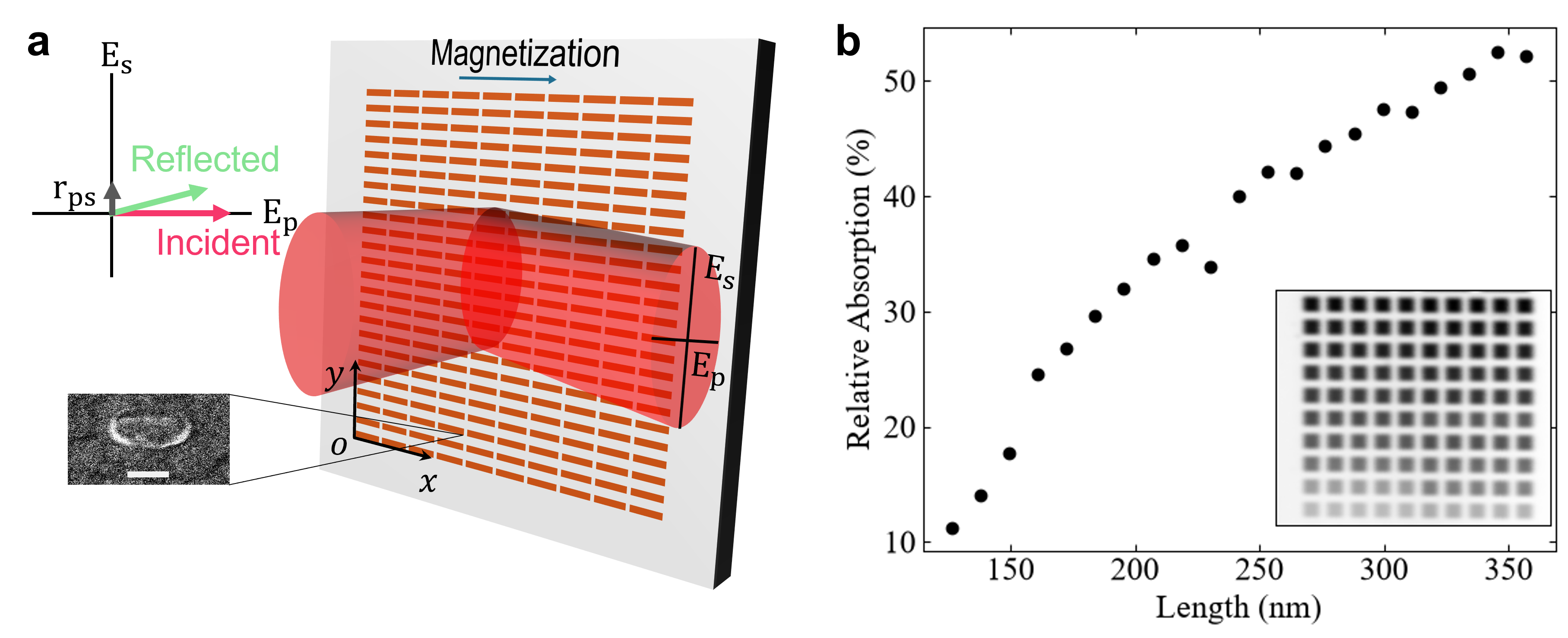}
\caption{ \textbf{MOKE measurement configuration and relative absorption.}  
\textbf{a}, Diagram of the longitudinal MOKE geometry. The incidence angle is 45\textdegree. 
Inset: Representative SEM image of the fabricated permalloy
nanobar. These are a separate set of samples from those in Figures~\ref{Figure:SEM_circlular_bar}
and ~\ref{Figure:circlular_bar_FTIR}, fabricated with the same four-layer cavity system as in
Figures~\ref{Figure:SEM_circlular_bar}b and c but with the nanobar width fixed at $W = 80$~nm. Scale
bar represents 100~nm.
\textbf{b}, Relative absorption (referenced to the bare cavity, see text) of 633 nm laser light for arrays with linearly increasing nanobar lengths, showing enhanced absorption with increasing length. Inset: Heatmap of absorption from scanning MOKE reflection measurements.}\label{Fig:MOKE measurement_Schematic}
\end{figure}

\begin{figure}
    \centering
    \includegraphics[width=\textwidth]{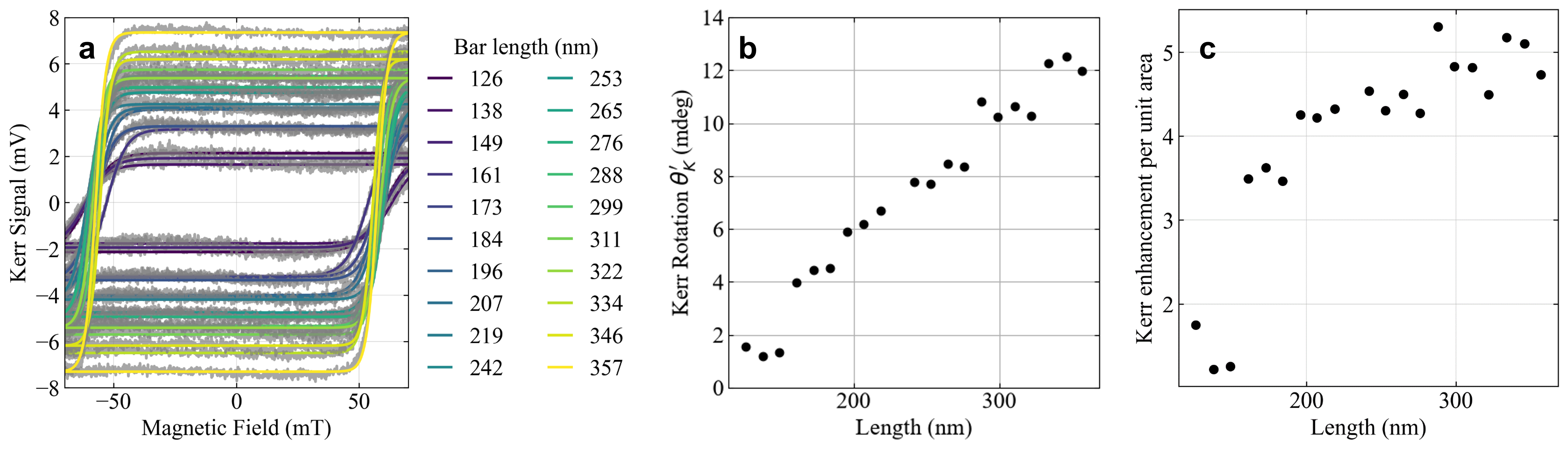}
    \caption{ \textbf{Results of the MOKE measurement.}  
    \textbf{a}, Hysteresis loops of the MOKE signal for nanobars of fixed width $W = 80$~nm and lengths ranging from 126 to 357~nm, illustrating the length-dependent magneto-optical response. The colour scale represents the nanobar length. \textbf{b}, Kerr rotation angle ($\theta^\prime_K$), obtained from the ratio of the change in saturated MOKE signals between the positively and negatively magnetized states ($\Delta I$) and the non-magneto-optical background signal ($I_{r0}$), plotted as a function of nanobar length. \textbf{c}, Kerr rotation per area enhancement factor, obtained by accounting for the area fill factor and normalized to the Kerr rotation of a continuous permalloy film. The reference film is an unpatterned 20~nm region of the same sample, deposited in the same evaporation step on the same interference cavity.}
    \label{Fig:naon_data}
\end{figure}

Through the design approach described, we have shown how to tune the interaction between light and a permalloy nanomagnet metasurface. In previous work, our motivation was to enhance all-optical switching of individual magnets within ASI \cite{stenning2023low}. As a consequence of enhanced magneto-optical interaction, we can also expect to enhance the MOKE in these systems. MOKE is a sensitive technique that probes a material's magnetization by analyzing the polarization change of reflected light upon interaction. While previous studies \cite{kim2020extreme} have used this concept to enhance polar MOKE signals in thin magnetic films, here we show that the enhancement is achievable for longitudinal MOKE in a metasurface of nanomagnets with in-plane magnetization. 

Figure \ref{Fig:MOKE measurement_Schematic} and Figure~\ref{Fig:naon_data} present the results of MOKE measurements on each full array of permalloy rectangular bars of varying lengths. Because the nanomagnets are magnetized in-plane, longitudinal MOKE requires non-normal incidence. Here we use a NanoMOKE2 system in a longitudinal MOKE configuration with 633 nm wavelength light incident at 45 degrees to the sample surface, as shown in Fig.~\ref{Fig:MOKE measurement_Schematic}a. The laser spot size is slightly smaller than the size of each array, so the measured response is averaged over the array. Figure \ref{Fig:MOKE measurement_Schematic}b shows the variation of the relative absorption as a function of bar nanomagnet length, evaluated by raster-scanning the illumination beam of the MOKE setup over the array, while recording the reflected intensity with the photodiode. The incident light had a wavelength of 633 nm and was linearly p-polarized. The relative absorption was inferred from the reflectance contrast between each patterned nanomagnet array and the adjacent bare cavity region (no permalloy). Since the bare cavity has a reflectivity below unity, this relative absorption is a lower bound on the absolute absorption of the array. We observe an increase in relative absorption as the magnet bar length increases to a maximum of 52\%. The corresponding absolute absorption remains lower than that observed under normal
incidence for the arrays of Figure~\ref{Figure:circlular_bar_FTIR}c and d; this reflects both the
$45^{\circ}$ angle of incidence and the smaller bar width ($W = 80$~nm)
used for the MOKE series, which reduces the fill factor. It nevertheless
remains substantial, demonstrating that strong light-matter interactions
persist under oblique illumination. The inset heatmap of Figure~\ref{Fig:MOKE measurement_Schematic}b further visualizes the variation in absorption from each of the measured magnetic metasurface arrays.

In longitudinal MOKE, p-polarised incident light is reflected with a small s-polarised component due to the magneto-optical Kerr effect, in addition to the dominant p-polarised component. The reflected electric field can be described as ${\bf E}_r= \left(r_{pp} \hat{{\bf p}}+r_{sp}\hat{{\bf s}}\right) E_0$, where $r_{pp}$ and $r_{sp}$ are the non-magneto-optical and magneto-optical reflection coefficients, respectively, while $\hat{{\bf p}}$ and $\hat{{\bf s}}$ are polarization vectors. The intensity of light transmitted through an analyser, set at a small offset angle $\theta_a$ from maximum extinction ($2^\circ$ in the current experiment), is given by $I_r = I_0|r_{pp} \sin \theta_a + r_{sp} \cos \theta_a|^2 \approx I_{r0}(1 + 2 \theta _K^\prime/\theta_a)$, for $|r_{sp}|<<|r_{pp}|$ and small $\theta_a$, where $I_{r0}\approx\theta_a^2 |r_{pp}|^2I_0$ is the reflected intensity without magnetic field and the Kerr rotation angle, $\theta_K=r_{sp}/r_{pp}=\theta _K^\prime + i\theta _K^{\prime\prime}$. Experimentally, we measure the intensity changes upon magnetization in forward ($I_{M_{+}}$) and reverse ($I_{M_{-}}$) directions to eliminate artifacts, where $\Delta I = I_{M_{+}}-I_{M_{-}} \approx 4I_{r0} \theta _K^\prime/\theta_a$.

Figure~\ref{Fig:naon_data}a shows the longitudinal MOKE hysteresis loops measured for nanobars with lengths ranging from 126 to 357~nm, with the different colors denoting the respective bar lengths. To quantify the length dependence of the MOKE response, the Kerr rotation angle at magnetic saturation was extracted from each hysteresis loop and plotted as a function of bar length in Figure~\ref{Fig:naon_data}\textbf{b}. Although the coercivity remains nearly independent of bar length, the magnitude of the normalized magneto-optical signal, $\Delta I/I_{r0}$, due to the Kerr polarization rotation, increases progressively with increasing bar length. The non-magneto-optical signal, $I_{r0}$, was background-corrected by subtracting the photodiode signal measured with the polarizer and analyzer in a crossed configuration, thereby accounting for contributions from the detector dark voltage and stray or scattered light. To provide a reference for evaluating the enhancement, we also measured
a Kerr rotation of 12.5~mdeg from a continuous permalloy reference region on the
same sample. This reference is a continuous 20~nm film deposited in the
same evaporation step, on the same interference structure, and measured
under identical conditions, so it differs from the arrays only in the
absence of nanoscale patterning. Figure~\ref{Fig:naon_data}\textbf{c} shows the Kerr rotation enhancement normalized by the permalloy nanobar area. Importantly, the optical absorption and MOKE response were measured using the same series of nanomagnet arrays fabricated on the same Al back-plane interference structure. The nanobar length was varied to tune the absorption, while the substrate, multilayer cavity, magnetic material, and fabrication conditions remained unchanged. After normalization by the permalloy nanobar area, the Kerr response increases with nanobar length, reaching approximately five times the Kerr rotation per unit area of the continuous reference film for the longest bars (Figure~\ref{Fig:naon_data}c). Because the substrate, cavity, magnetic material, film thickness and measurement conditions are common to the entire series, this cannot be attributed simply to a larger amount of permalloy within the illuminated spot. This comparison shows that nanomagnets with stronger absorption generate a larger MOKE response per unit area of magnetic material.

The enhancement of MOKE polarization rotation, $\theta_K\approx r_{sp}/r_{pp}$, with bar length can be attributed to both an increased magneto-optical cross-reflection coefficient, $r_{sp}$, and reduced specular reflection, $r_{pp}$, in the interference design. Near-field coupling between adjacent nanostructures is expected to be weak in the present geometry because of the relatively large inter-structure spacing. The enhanced MOKE is therefore primarily attributed to interference-enhanced absorption in the multilayer cavity system, while collective near-field coupling is expected to play a secondary role.

We note that the enhancement is expressed per unit area of magnetic material. In absolute terms the largest arrays reach a Kerr rotation comparable to that of the continuous reference film while covering only about a fifth of the surface, so the per-area figure is the appropriate measure of how efficiently light is coupled into the magnetic material. The absolute signal is bounded by the fill factor and could be raised further by denser arrays. We also note that the reference film sits on the same interference cavity as the arrays, so the quoted enhancement is referenced to an already cavity-enhanced baseline.

It should be noted that the present MOKE measurements were performed at a fixed wavelength of 633 nm because the available magneto-optical setup employs a fixed-wavelength laser source. Importantly, 633 nm corresponds to the design wavelength of the interference structure. The MOKE measurement therefore directly probes the Kerr response under the interference-assisted absorption condition investigated in this work. In addition, the extracted Kerr rotation angles and enhancement factors should be regarded as semi-quantitative estimates. The analysis assumes purely p-polarized incident light, an ideal linear analyzer, and fully coherent specular reflection, conditions that are not strictly realized because of optical imperfections and alignment uncertainties. Comparison with literature magneto-optical constants is further complicated by the sensitivity of patterned permalloy to fabrication conditions.

\section{Conclusions}

We have shown that multilayer interference engineering offers an effective and robust strategy for dramatically enhancing optical absorption in nanoscale magnetic metasurfaces. By integrating permalloy nanomagnet arrays with a thin dielectric spacer and a reflective metallic back-plane, we achieve a metasurface environment in which geometric parameters—including nanomagnet shape, fill factor, and cavity thickness—determine the interference conditions for incident light. Tuning these parameters allows the system to reach destructive interference in reflection, enabling near-unity absorption concentrated within the magnetic nanostructures.

This absorption enhancement is accompanied by stronger MOKE signals, driven by both increased light-nanomagnet interaction and reduced reflected background. The 500--1200 nm range investigated here follows from the interference condition designed at an operating wavelength of 633 nm, together with the optical properties of the constituent materials and the response range of our FTIR setup. Importantly, the underlying mechanism is general and can be extended to other wavelength regimes through appropriate material selection and geometric scaling. The demonstrated strategy addresses the limited optical interaction volume of ultrathin magnetic nanostructures and provides a pathway toward compact and efficient opto-magnetic platforms. More broadly, our approach is applicable to a wide range of photonic, plasmonic, and metasurface-based structures where strong absorption in ultra-thin functional layers is desired. These findings open new directions in optically driven magnetism, magnetic metasurface engineering, and light-enabled spintronic technologies.

\section{Methods}

\subsection*{Optical simulation details}
 Simulations of the linear optical response of the nanostructures were performed using the finite-difference time-domain (FDTD) method. Linearly polarized plane waves of varying wavelength and polarization were injected from the air side onto the periodic array; the polarization used in each case is indicated in the corresponding figure. Periodic boundary conditions were applied in the x- and y-directions, and perfectly matched layers were used in the z-direction to absorb outgoing light with minimal back-reflection. The optical constants used in the FDTD simulations were taken from experimentally measured wavelength-dependent datasets. The optical constants of permalloy were taken from Ref. \citenum{tikuivsis2017optical}, Au from the Johnson and Christy dataset~\cite{johnson1972optical}, and SiO$_2$, Al$_2$O$_3$, Si, and Al from Palik~\cite{palik1998handbook}. The corresponding complex dielectric functions are plotted in the Supporting Information (Figure~S3). The wavelength-dependent optical properties of these materials determine the interference, absorption, and field confinement characteristics of the metasurface system within the investigated spectral range. For the ferromagnetic permalloy, although the dielectric function becomes tensorial in the magnetized state, only the diagonal complex permittivity was used in the FDTD simulations, as the off-diagonal terms are much smaller and are expected to have a negligible effect on the absorption features considered here, consistent with the measurements in Supporting Information Figure S2.

\subsection*{Experimental methods}
Samples were fabricated via electron-beam lithography and thermal evaporation. Two back-plane variants were used: Au and Al. For the Au back-plane, a Cr(2~nm)/Au(250~nm)/Cr(2~nm) stack was thermally evaporated onto a silicon wafer, the Cr layers improving adhesion of Au to the substrate and of the silica to the Au. For the Al back-plane, a 250~nm Al film was thermally evaporated onto a silicon wafer, thick enough that transmission is negligible across 500--1200~nm. A SiO$_2$ spacer of the stated nominal thickness was then deposited by sputtering in an Angstrom Engineering AMOD system, with the thickness confirmed by ellipsometry. All samples reported here use SiO$_2$; Al$_2$O$_3$ is included in Figure~\ref{Fig:Schematic}a as an equivalent low-loss alternative for the spacer layer. Then, the substrate was spin-coated with a bi-layer 495K / 950K poly(methyl methacrylate)  (PMMA) resist layer and baked at 180~$^{\circ}$C for 5 min. The nanostructures were subsequently defined by electron-beam exposure, followed by development in a standard 1:3 methyl isobutyl ketone/isopropanol (MIBK/IPA) solution (for 40 s). A 20 nm permalloy (Ni$_{81}$Fe$_{19}$) film was thermally evaporated at a base pressure of $2 \times 10^{-6}$~mbar,  and the lift-off process was performed in acetone as a final step.

\subsection*{MOKE experiments}
The longitudinal MOKE measurements were performed using a Durham Magneto-Optics NanoMOKE2 system at room temperature. A 633 nm diode laser was focused to a $\sim$20 $\mu m$ spot diameter onto the nanomagnet array and an external magnetic field was applied in-plane, parallel to the long axis of the magnetic bars. To compare the different array dimensions the incident laser polarization was kept constant manually at an offset of $2^\circ$ from the extinction angle defined by the polariser-analyser alignment.

Hysteresis loops were measured under a quasistatic sinusoidal magnetic field generated by a quadrupole electromagnet operating at 10 Hz. Signals were averaged over 1000 field cycles. The MOKE signal was normalized by linear background subtraction.

\subsection*{Fourier transform infrared measurements}
FTIR spectroscopy was used to characterize the linear optical response of the fabricated nanomagnet arrays. Spectra were collected with a Bruker Hyperion 2000 FTIR microscope equipped with a 36×, NA = 0.5 reflective objective. Reflection spectra were obtained by normalizing the signal from the patterned array region to a reference spectrum recorded from a silver mirror located close to the array on the same sample. Measurements were made over 500–1100 nm and 1100–1200 nm using a silicon detector and a Peltier-cooled InGaAs detector, respectively.

\subsection*{Author contributions}
X.X., A.V., J.C.G., R.F.O., and W.R.B. conceived the work. X.X. drafted the manuscript other than the description of the MOKE characterization and enhancement with the help from R.F.O.. A.V. drafted the description of the MOKE section with the help from T.Z.. A.V., K.D.S., and J.C.G. fabricated the samples. X.X. and H.H. performed FTIR measurements. X.X. performed modeling and numerical simulations for the optical properties of the structures. T.Z., A.V. performed the MOKE measurements and analysis with the help from J.C.G., D.S.B., and W.R.B.. All authors contributed to the revision stages.

\subsection*{Competing interests}
The authors declare no conflict of interest.

\begin{acknowledgement}

\end{acknowledgement}

\bibliography{achemso-demo}

\end{document}